\documentclass[10pt,preprintnumbers,aps,amssymb,nofootinbib,amsmath,superscriptaddress,prd,notitlepage]
{revtex4-2}
\usepackage{epsfig,epsf}
\usepackage{comment}
\usepackage{bm} 
\usepackage{color} 
\usepackage{slashed}
\usepackage{relsize}	
\usepackage{soul} 
\usepackage{hyperref}
\usepackage{tensor} 
\usepackage{yfonts} 
\newcommand{\beq}{\begin{equation}}
\newcommand{\beql}[1]{\begin{equation}\label{#1}}
\newcommand{\eeq}{\end{equation}}
\def\bal#1\gal{\begin{align}#1\end{align}}
\newcommand{\ball}[1]{\bal\label{#1}}
\newcommand{\eq}[1]{(\ref{#1})}
\newcommand{\fig}[1]{Fig.~\ref{#1}}
\renewcommand{\sec}[1]{Sec.~\ref{#1}}
\DeclareMathOperator{\im}{\mathrm{Im}}

\renewcommand{\b}[1]{{\bm #1}} 
\newcommand{\unit}[1]{\hat {{\bm #1}}} 

\newcommand{\aver}[1]{\left\langle #1 \right\rangle}

\newcommand{\mg}[1]{|{\bm{#1}}|}
\begin{document}


\title{Photon emission from rotating plasmas: a generalized McLerran-Toimela formula and the onset of superradiance}

\author{Kirill Tuchin}

\affiliation{
Department of Physics and Astronomy, Iowa State University, Ames, Iowa, 50011, USA}

\date{\today}


\begin{abstract}

The photon emission and absorption spectra of a plasma rotating with constant angular velocity $\Omega$ are derived in terms of the spectral function of the current--current correlator. In contrast to the non-rotating case, the leading contribution arises already at one-loop order. The photon emission spectrum and its elliptic flow are calculated for a rotating quark--gluon plasma and compared with the leading two-loop result for a non-rotating plasma. The rotating plasma is found to emit significantly more soft photons than the non-rotating one. 

An analysis of the emission and absorption of cylindrical waves indicates that photons with energy $\omega$ and azimuthal quantum number $m$ satisfying  $\omega<m\Omega$ are emitted at a higher rate than they are absorbed, thereby exhibiting the phenomenon of superradiance. It is further argued that these superradiant modes induce an instability of the magnetic field that is generated concurrently with the plasma in relativistic heavy-ion collisions.

\end{abstract}

\maketitle

\section{Introduction}\label{sec:a}

Plasmas emit electromagnetic radiation, which serves as a valuable diagnostic tool for studying plasma properties. While plasmas often rotate, the rotation is usually too slow to have a significant impact on radiation compared to other sources.  A notable exception is the quark-gluon plasma, where measurements have revealed substantial vorticity \cite{STAR:2017ckg, STAR:2018gyt, ALICE:2019aid, Becattini:2024uha, Niida:2024ntm}. The rotation of quark-gluon plasma can significantly increase the intensity of its electromagnetic radiation \cite{Buzzegoli:2023yut}. This paper computes the electromagnetic radiation of rotating plasmas, with particular focus on the quark-gluon plasma.

Previous studies of electromagnetic radiation from rotating plasmas  focused on synchrotron radiation, which requires the presence of a strong magnetic field \cite{Buzzegoli:2022dhw,Buzzegoli:2023vne,Buzzegoli:2025qfl,Buzzegoli:2026sji}. In contrast, this paper completely ignores the effect of the magnetic field on the plasma. 

The primary goal of the present work is to establish a general relationship between the photon spectrum and the photon polarization tensor in a rotating plasma. This goal is accomplished by Eq.~\eq{d13}, the main result of this paper. The analogous relation for non-rotating plasmas, the McLerran-Toimela formula \cite{Weldon:1983jn,McLerran:1984ay,Gale:1990pn}, has become a cornerstone for phenomenological applications in relativistic heavy-ion collisions and astrophysical plasmas. 

The derivation of \eq{d13}, which relates the photon spectrum to the imaginary part of the retarded photon polarization tensor, is presented in Sections~\ref{sec:b}--\ref{sec:d}. \sec{sec:b} derives the photon emission and absorption probabilities in terms of the Wightman functions, following the standard procedure. In view of the axial symmetry of a rotating plasma, the photon wave functions are chosen in the cylindrical wave basis in \sec{sec:c}. The cylindrical waves  are characterized by transverse momentum $k_\bot$, longitudinal momentum $k_z$ and azimuthal quantum number $m$; alternatively, $k_\bot$ and $k_z$ can be replaced by photon's energy and emission angle $\theta$ with respect to the rotation axis, taken to be the $z$-axis. \sec{sec:d}  derives the final result and discusses its dependence on the plasma's extent in the radial direction.

Sections \ref{sec:f} and \ref{sec:g} consider two applications of \eq{d13} in the context of relativistic heavy-ion phenomenology. The first application concerns photon emission by a rotating quark-gluon plasma. The leading contribution to photon emission by a rotating plasma starts at the one-loop order. In contrast, in a non-rotating quark-gluon plasma the leading contribution to the photon spectrum appears at order $ \alpha\alpha_s $ (two-loop order): rotation supplies additional energy and momentum, opening the phase space and rendering the one-loop contribution finite. This one-loop term is evaluated in \sec{sec:f}, yielding the photon spectrum given in Eq.~\eq{f5}. The resulting spectrum is shown in \fig{fig:spect}, together with the elliptic flow coefficient $v_2$. At $\omega\gtrsim 1$~GeV, the rotating-plasma result is of the same order of magnitude as the leading non-rotating two-loop result (A4). However, it exhibits a significant enhancement in the infrared region.  The elliptic flow $v_2$ is positive above  $\omega\simeq 1.5$~GeV and negative otherwise.

The second application concerns superradiance in rotating systems. Any rotating dissipative system, such as plasma, exhibits superradiance --- the spontaneous emission of electromagnetic waves with energies satisfying $\omega<m\Omega$ \cite{Zeldovich:1971ffh,Zeldovich:1972zqp,Misner:1972kx}. While often discussed in the context of black holes, superradiance is a generic feature of rotating systems \cite{Bekenstein:1998nt,Endlich:2016jgc}. However, only a small fraction of the total electromagnetic radiation produced by a rotating plasma arises from this superradiant mechanism. 

The phenomenological significance of superradiance lies in its ability to amplify infrared modes of electromagnetic field incident on the plasma. It is well known that a strong magnetic field accompanies quark-gluon plasma in relativistic heavy-ion collisions \cite{Kharzeev:2007jp,Skokov:2009qp,Voronyuk:2011jd,Bzdak:2011yy,Bloczynski:2012en,Deng:2012pc,Tuchin:2013apa,Zakharov:2014dia}. Although the strength of this magnetic field decreases significantly over time, it remains substantial throughout the plasma's lifetime \cite{Stewart:2021mjz}. A goal of this study is to investigate the superradiant amplification of this magnetic field by the quark-gluon plasma.  It is shown in \sec{sec:g} that plasma superradiance drives an instability of the magnetic field. For a plasma with a lifetime of 10~fm, the phenomenologically significant domain of photon energies where this instability arises is below $\sim 10$~MeV. The growth and damping rates of  various modes are displayed in \fig{fig:super}.

A discussion of the results and conclusions is presented in Sec.~\ref{sec:k}.

The natural units $\hbar=c=k_B=1$, $\alpha= e^2/4\pi$ are employed throughout. The photon's momentum, energy,  and four-momentum are denoted $\b k$,  $\omega$, and $k$, respectively, and the $z$-axis is taken as the rotation axis. The QCD version of equations is obtained by replacing $e^2\to e^2N_c \sum_f q_f^2 = 5e^2/3$ for two flavors. 

\section{Photon absorption and emission by a rotating system}\label{sec:b}

Consider a homogeneous plasma rotating with constant angular velocity $\Omega$ about the $z$-axis. It is assumed that rotation does not alter the local properties of the plasma, which amounts to neglecting non-inertial forces in the comoving frame;\footnote{The leading non-inertial specific force due to rotation is $\b v\times \b \Omega$, while the typical specific shear viscous force is $\eta \Delta \b v/w$, where $\eta$ is the viscosity and $w$ is the enthalpy density. The ratio of the former to the latter is of the order of $\lambda=(L\Omega) (LT) (s/\eta)$, where $L$ is a length scale over which the velocity field varies. At an early stage after heavy-ion collsion $T=350$~MeV, $\eta/s\approx 0.6$ \cite{Niemi:2012ry}.  Taking $\Omega=0.05$~fm$^{-1}$ and $L=1$~fm \cite{Becattini:2015ska}, we find that $\lambda\approx 0.1$. At later stages, the value of $\lambda$ progressively increases. 
}
 a recent analysis of this issue is given in \cite{Facundes:2025xcc}. It is further assumed that the light cylinder, located at radial distance $\Omega^{-1}$, lies outside the plasma, so that the entire plasma is a causally connected manifold.  

Absorption of a photon excites the plasma from an initial state $|i\rangle$ to a new state $|f\rangle$. The corresponding amplitude is given by a matrix element of the $S$-matrix:
\ball{b1}
\langle f;0|S|i;\gamma\rangle= i\int d^4x\langle f;0|R^{kl}(t)j^l(x)A^k(x)|i;\gamma\rangle = i\int d^4x\langle f;0|j^l(x)(R^{-1})^{lk}(t)A^k(x)|i;\gamma\rangle\,,
\gal
where $j^\mu(x)= e\bar \psi \gamma^\mu \psi$ is the electric current density operator in the plasma rest frame, $R^{kl}(t)$ is the rotation matrix about the $z$-axis, $A^\mu(x)$ is the electromagnetic potential operator in the radiation gauge. 
The absorption probability is given by 
\bal
w_\text{abs}&= \sum_f\left|\langle f;0|S|i;\gamma\rangle \right|^2 d\Gamma\nonumber 
\\
&=\int d^4x \int d^4x' (R^{-1})^{nm}(t') (R^{-1})^{lk}(t) \sum_f   \langle i|j^n(x')|f\rangle  \langle f|j^l(x)|i\rangle 
\langle 0| A^k(x)|\gamma\rangle \langle \gamma| A^m(x')|0\rangle d\Gamma \nonumber 
\\
&= \int d^4x \int d^4x' (R^{-1})^{nm}(t') (R^{-1})^{lk}(t)    \langle i|j^n(x')j^l(x)|i\rangle 
\langle 0| A^k(x)|\gamma\rangle \langle \gamma| A^m(x')|0\rangle d\Gamma\,. \label{b5}
\gal
where $d\Gamma$ denotes  the element of the photon's phase space. Define the Wightman correlator 
\ball{b7}
i\Pi^{nl}_+(x',x)=   \langle i|j^n(x')j^l(x)|i\rangle \,,
\gal
where the conventions of Kapusta and Gale \cite{Kapusta:2006pm} are used. 

In its rest frame, a plasma is translationally invariant if its boundary can be ignored. However, this premise is not generally valid for a rotating plasma, as its boundary is essential due to causality \cite{Buzzegoli:2023vne,Ambrus:2015lfr}, as discussed in detail in the next two sections. Nevertheless, the impact of the boundary on the Wightman correlator 
can be neglected at high temperatures. A rotating plasma of radial extent $\mathcal{R}$ at temperature $T$ can be assumed to be locally translationally invariant if $\mathcal{R}T\gg 1$. This also allows treating the quasi-particle spectra as continuous. Assuming that this condition holds, the  Fourier image of \eq{b7} reads
\ball{b9}
i\Pi^{nl}_+(x',x)= \int \frac{d^4q}{(2\pi)^4}e^{iq\cdot (x-x')}i\Pi^{nl}_+(q)\,.
\gal
The finite-size effects and physical boundaries in relativistic rotating matter were previously studied in \cite{Ambrus:2015lfr,Ebihara:2016fwa,Chernodub:2016kxh,Chernodub:2017ref,Buzzegoli:2022dhw,Buzzegoli:2024nzd}.

The Wightman function in momentum space is real and positive: $i\Pi^{nl}_+(q)\ge 0$.   Substituting \eq{b7} and \eq{b9} into \eq{b5}, the absorption probability can be cast in the form
\ball{b11}
w_\text{abs}=\int \frac{d^4q}{(2\pi)^4} i\Pi^{nl}_+(q) \psi^l(q) \psi^{n*}(q) d\Gamma\,,
\gal
where the auxiliary vector $\b\psi$ is defined as 
\ball{b12}
 \psi^l (q) = \int d^4 x e^{iq\cdot x}(R^{-1})^{lk}(t) \langle 0| A^k(x)|\gamma\rangle \,.
\gal

In thermal equilibrium at temperature $T=\beta^{-1}$, the absorption probability can be expressed in terms of the imaginary part of the retarded Green's function via the relation
\ball{b13}
i\Pi^{nl}_+(q)= -2[1+n_B(q^0)]\im\Pi^{nl}_R(q)\,,
\gal
where  $n_B(q^0)= (e^{\beta q^0}-1)^{-1}$ is the Bose distribution. Hence, \eq{b11} becomes
\ball{b14}
w_\text{abs}=-2\int \frac{d^4q}{(2\pi)^4} [1+n_B(q^0)]\im\Pi^{nl}_R(q) \psi^l(q) \psi^{n*}(q) d\Gamma\,.
\gal

The emission probability can be computed in a similar way:
\bal
w_\text{em}&= \sum_f\left|\langle f;\gamma|S|i;0\rangle \right|^2 d\Gamma
= -\int \frac{d^4q}{(2\pi)^4} i\Pi^{nl}_-(q) \psi^l (q)\psi^{n*}(q)d\Gamma\,, \label{b15}
\gal
where 
\ball{b17}
i\Pi^{nl}_-(x',x)=  - \langle i|j^l(x)j^n(x')|i\rangle \,.
\gal
The function $i\Pi^{nl}_-(q)$ is real and negative. Using the relation
\ball{b18}
i\Pi^{nl}_-(q)= 2n_B(q^0)\im\Pi^{nl}_R(q)\,,
\gal
Eq.~\eq{b15} becomes
\bal
w_\text{em}&= -2\int \frac{d^4q}{(2\pi)^4} n_B(q^0)\im\Pi^{nl}_R(q) \psi^l(q) \psi^{n*}(q)d\Gamma\,, \label{b19}
\gal

The difference between the emission and absorption probabilities is 
\bal
 w&= w_\text{em}-w_\text{abs}=2\int \frac{d^4q}{(2\pi)^4} \im\Pi^{nl}_R(q) \psi^l(q) \psi^{n*}(q) d\Gamma\,.\label{b20}
\gal

No choice of electromagnetic wave basis has been made so far. For a non-rotating plasma, it is convenient to use the plane-wave basis — that is, states of definite photon momentum $\b k$. It is shown in Appendix~\ref{app:A} that in this case \eq{b20} reduces to the well-known result \eq{z5}.

\section{Photon's cylindrical waves}\label{sec:c}

The boundary of a rotating system cannot be ignored, as it can be in the non-rotating case, and requires careful consideration.  Suppose the rotating plasma is bounded by a cylindrical surface of radius $r=\mathcal{R}$ and height (extent in the $z$-direction) $L$. In view of the axial symmetry, it is expedient to choose the cylindrical wave basis for the electromagnetic field.  

Cylindrical waves were previously used by Buzzegoli et al.\ to compute the synchrotron radiation of photons from a rotating plasma \cite{Buzzegoli:2023vne,Buzzegoli:2022dhw}. The wavefunction of a photon with longitudinal momentum $k_z$,  transverse momentum magnitude $k_\bot$,  azimuthal quantum number $m$, and  polarization $\lambda$, is given by 
\ball{c1}
\langle 0| \b A(x)|\gamma(k_\bot,k_z,m,\lambda)\rangle = \frac{1}{\sqrt{2\omega V}}
\b \Phi_{k_\bot,k_z,m,\lambda}(\b x)e^{-i\omega t }\,,
\gal
where 
\begin{subequations}\label{cc1}
\ball{c3}
\b \Phi_{k_\bot,k_z,m,\lambda}(\b x)= \frac{|\b k|}{k_\bot}\frac{1}{\sqrt{2}}\left( \lambda \b T_{k_\bot,k_z,m,\lambda}(\b x)+ \b P_{k_\bot,k_z,m,\lambda}(\b x)\right)\,,
\gal
with the toroidal  and poloidal  waves given by
 \bal
\b T_{k_\bot,k_z,m,\lambda}(\b x)&= -\frac{1}{|\b k|}(\unit z\times \b \nabla) u(\b x)\,, \label{c5}\\
\b P_{k_\bot,k_z,m,\lambda}(\b x)&= \frac{1}{|\b k|}\b \nabla \times \b T_{k_\bot,k_z,m,\lambda}(\b x)\,,\label{c6}
\gal
where 
\ball{c7}
u(\b x)=J_m(k_\bot r)e^{i(k_zz+m\phi)}
\gal
\end{subequations}
is a solution of the Helmholtz equation  in cylindrical coordinates $\b x = (\phi,r,z)$. These waves satisfy the orthogonality condition
\ball{c8}
\int \b \Phi_{k_\bot,k_z,m,\lambda}(\b x)\cdot  \b \Phi_{k'_\bot,k'_z,m',\lambda'}(\b x) d^3x = (2\pi)^2\delta_{mm'}\delta_{\lambda\lambda'}\frac{\delta(k_\bot-k_\bot')}{k_\bot}\delta(k_z-k_z')\,.
\gal

The angular momentum operator $L_z= -i\partial_\phi$ commutes with the operators $\unit z\times \b \nabla=\unit \phi\partial_r-\unit r   r^{-1}\partial_\phi$ and curl. Hence the functions \eq{cc1} are chosen as simultaneous eigenfunctions of $L_z$ and curl:
\bal
L_z\b \Phi_{k_\bot,k_z,m,\lambda}(\b x)&= m\b \Phi_{k_\bot,k_z,m,\lambda}(\b x)\,.\label{c9}\\
\b \nabla \times \b \Phi_{k_\bot,k_z,m,\lambda}(\b x) &= \lambda|\b k| \b\Phi_{k_\bot,k_z,m,\lambda}(\b x)\,.\label{c10}
\gal

The following integral emerges in the next section:
\ball{c11.A}
\b \Phi_{k_\bot,k_z,m,\lambda}^\mathcal{R}(\b q)&= \int_{r\le \mathcal{R}} e^{-i\b q\cdot \b x}\b \Phi_{k_\bot,k_z,m,\lambda}(\b x)d^3x
= 
-\frac{(2\pi)^2}{\sqrt{2}}(-i)^{m+1}\frac{1}{k_\bot}\b I(\b q) f(q_\bot) \delta(q_z-k_z)\,,
\gal
where 
\ball{c13.A}
\b I(\b q) = \lambda \b q\times \unit z+ \frac{i}{k}\b q\times(\b q\times \unit z)\,,
\gal
and 
\ball{c14}
f_{k_\bot,m}(q_\bot)= \frac{\mathcal{R}}{k_\bot^2-q_\bot^2}\left[ q_\bot J_{m-1}(q_\bot \mathcal{R})J_m(k_\bot \mathcal{R})- k_\bot J_{m-1}(k_\bot \mathcal{R})J_m(q_\bot \mathcal{R})\right]\,.
\gal
When $\mathcal{R}\to \infty$, $f(q_\bot)\to \delta(q_\bot-k_\bot)/k_\bot$, a consequence of the closure equation of the Bessel functions. However, this limit cannot be taken in a rotating system, because the sum over the azimuthal quantum number $m$ in the photon spectrum would then diverge. This occurs because the contribution to the sum over $m$ from photons with $|m|\gg k_\bot \mathcal{R}$  decreases as $e^{-4|m|\log |m|}$, a suppression that is lost if the plasma radius is taken to infinity.

The phase space of cylindrical waves with azimuthal quantum number $m$ is given by \cite{Buzzegoli:2023vne,Buzzegoli:2022dhw}
\ball{c15}
d\Gamma = \frac{dk_\bot k_\bot dk_z \pi \mathcal{R}^2L}{(2\pi)^2}\,.
\gal

\section{Emission and absorption by a rotating cylinder}\label{sec:d}

Given a photon in a cylindrical wave state with longitudinal momentum $k_z$, transverse momentum magnitude $k_\bot$,  azimuthal quantum number $m$, and polarization $\lambda$, the vector $\b\psi$ defined in \eq{b12} is evaluated using  \eq{c1},\eq{c9} and \eq{c11.A}:
\bal
 \b\psi (q) = \int_{r\le \mathcal{R}}  e^{iq\cdot x}R^{-1}(t) \langle 0|\b  A(x)|\gamma\rangle d^4 x= \int_{r\le \mathcal{R}}   e^{iq\cdot x+im\Omega t}\langle 0|\b  A(x)|\gamma\rangle d^4 x \nonumber\\
 = \frac{2\pi}{\sqrt{2\omega V}}\delta\left(q^0-\omega+m\Omega\right)\b \Phi_{k_\bot,k_z,m,\lambda}^\mathcal{R}(\b q)\,, \label{d2}
\gal
where $R= \exp\{-iL_z \Omega t\}$ is the rotation operator. Substituting \eq{d2} and \eq{c15} into \eq{b20} yields 
\ball{d4}
\dot w = \sum_{m,\lambda}\int\frac{d^4q}{(2\pi)^4}2\im \Pi_R^{ij}(q^0,\b q)\frac{(2\pi)^4}{2k_\bot^2}\delta(k_z-q_z)f^2_{k_\bot,m}(q_\bot)\frac{L}{2\pi}\frac{I^iI^{*j}}{2\omega V}2\pi \delta\left(q^0-\omega+m\Omega\right)
\frac{dk_\bot k_\bot dk_z \pi \mathcal{R}^2L}{(2\pi)^2}\,,
\gal
where the overdot on $w$ indicates the time derivative. The squares of the delta functions are interpreted in the usual way: one of the energy delta functions  produces $t/(2\pi)$, where $t$ is the observation time, and one of the longitudinal momentum delta functions produces $L/(2\pi)$. 

Since the plasma is assumed isotropic in its rest frame, the retarded photon correlator can be decomposed as \cite{Kapusta:2006pm}
\ball{d7}
\Pi^{ij}_R(q^0,\b q)= \Pi_T(q^0,|\b q|)\left( \delta^{ij}-\frac{q^iq^j}{\b q^2}\right)+ \Pi_L(q^0,|\b q|)\frac{q^iq^j}{\b q^2}\frac{(q^0)^2}{q^2}\,.
\gal
Furthermore, it follows from \eq{c13.A}  that $\b I\cdot \b q=0$. Therefore, 
\ball{d9}
\im\Pi^{ij}_R(q^0,\b q)I^iI^{*j}= \im\Pi_T(q^0,|\b q|)|\b I|^2= \im\Pi_T(q^0,|\b q|)q_\bot^2\left( 1+\frac{\b q^2}{\b k^2}\right)\,.
\gal
Collecting all formulas above, and recalling that $|\b k|=\omega$ for the emitted photon, furnishes the final result for the balance of the emission and absorption rates of a photon with given $k_z$ and  $k_\bot$:
\ball{d13}
\frac{d\dot w}{2\pi k_\bot dk_\bot dk_z V}= \sum_{m=-\infty}^\infty\int_0^\infty \frac{dq_\bot^2}{(4\pi)^2 \omega}\frac{q_\bot^2}{k_\bot^2}\left( 1+\frac{k_z^2+ q_\bot^2}{\omega^2}\right)\im\Pi_T\left(\omega-m\Omega, \sqrt{k_z^2+q_\bot^2} \right) f^2_{k_\bot,m}(q_\bot) \frac{1}{\pi\mathcal{R}^2}\,,
\gal
where $\omega=\sqrt{k_z^2+k_\bot^2}$. $ \im\Pi_T(q^0,|\b q|) <0$ when $q^0>0$ and $ \im\Pi_T(q^0,|\b q|) >0$ when $q^0<0$. This indicates that when a photon's azimuthal quantum number $m$ and energy $\omega$ satisfy the condition $\omega<m\Omega$, the rotating system emits more photons than it absorbs --- the phenomenon known as superradiance.

The effect of rotation on the photon spectrum in \eq{d13} is two-fold. Firstly, it shifts the photon energy by the amount $-m\Omega$. Secondly, it constrains photon emission to a cylindrical domain with a radius of $\mathcal{R}$. Although the particular form of the boundary is model dependent, the existence of a boundary is a physical requirement. This is because the velocity of the plasma periphery, which is proportional to the radial distance from the rotation axis, cannot exceed the speed of light. For a cylinder this condition is $\mathcal{R}\Omega<1$. It is therefore impossible to take the limit $\mathcal{R}\to \infty$ while keeping $\Omega$ fixed: a rotating plasma must be bound.  

The non-rotating limit is obtained by first taking  $\Omega\to 0$ and then  $\mathcal{R}\to \infty$. As discussed at the end of  \sec{sec:c}, the latter limit converts the function $f_{k_\bot,m}(q_\bot)$ into a delta-function, setting $q_\bot = k_\bot$, so that \eq{d13} reduces to \eq{z5}.

As mentioned earlier, superradiant photons satisfy the condition $\omega<m\Omega$. Since in practice $\omega\gg \Omega$, it is instructive to examine the behavior of the photon emission rate at large values of $|m|$ before performing the summation over $m$. Eq.~\eq{c14} shows that the photon emission amplitude is proportional to the Bessel function $J_m(k_\bot r)$. At large values of $|m|\gg k_\bot r$, the Bessel function is suppressed by the factor $|m|^{-1/2}(k_\bot r/|m|)^{|m|}$. The main contribution to the sum therefore stems from $|m|\lesssim k_\bot \mathcal{R}$. An equivalent way to see this is to note that a photon with a given $m$ and $k_\bot$ is emitted within the distance $r\sim m/k_\bot$ from the rotation axis; since $r$ must lie within the plasma, it again follows that $|m|\lesssim k_\bot\mathcal{R}<\omega \mathcal{R}$. On the other hand, the velocity of the cylinder periphery cannot exceed unity: $\Omega\mathcal{R}< 1$, implying  $|m|\lesssim \omega/\Omega$ --- a condition in tension with that  for superradiance. Hence, the domain of superradiance corresponds to the values of  $m$ that make only a subleading contribution to the total emission rate. These modes become leading, however, in the reflection problem, discussed further in \sec{sec:g}.

There is a particular case where taking the limit of large $\mathcal{R}$ first is justified: the contribution of the modes with 
 $|m|\ll k_\bot \mathcal{R}$. The photon spectrum for a given azimuthal quantum number $m$ then reduces to a particularly simple form:
 \ball{d14}
\frac{d\dot w}{2\pi k_\bot dk_\bot dk_z V}= \frac{1}{2\pi^4}\frac{1}{\mathcal{R}k_\bot \omega} \im\Pi_T(\omega-m\Omega,\omega )\,, \qquad m\ll k_\bot \mathcal{R}\,.
\gal
However, since $\im\Pi_T(q^0,\mg q)$ is not bound at large values of $|q^0|$, the sum over $m$ in the right-hand side of \eq{d14} would diverge. A simple fix to this problem, imposing a cutoff $|m|\le m_\text{max}=k_\bot \mathcal{R}$, is not satisfactory as the resulting spectrum would have a strong dependence on this cutoff. Therefore, as far as the total, summed over $m$, spectrum is concerned, \eq{d13} gives the most accurate approximation.

The next two sections consider two applications of the main result \eq{d13}, as advertised in the Introduction.

\section{Photon production by rotating quark-gluon plasma at one-loop order}\label{sec:f}

In relativistic heavy-ion collisions, one is interested in the production of photons from the quark-gluon plasma. Photon radiation arises from thermal fluctuations, since a perfectly homogenous plasma does not radiate at $T=0$. Assuming there is no incident electromagnetic field, the relevant observable is photon emission probability.

Consider a cylindrical domain of the quark-gluon plasma rotating with constant angular velocity $\Omega$. The photon emission spectrum can be computed using \eq{b15}. Following the same steps that led from \eq{d2} to \eq{d13} yields 
\ball{f5}
\frac{d\dot w_\text{em}}{2\pi k_\bot dk_\bot dk_z V}=-\sum_{m}\int_0^\infty \frac{dq_\bot^2}{(4\pi)^2 \omega}\frac{q_\bot^2}{k_\bot^2}\left( 1+\frac{k_z^2+ q_\bot^2}{\omega^2}\right)n_B(\omega-m\Omega)\im\Pi_T\left(\omega-m\Omega, \sqrt{k_z^2+q_\bot^2} \right)  \frac{f^2_{k_\bot,m}(q_\bot)}{\pi\mathcal{R}^2}\,.
\gal
Owing to the identity $n_B(-x)= -1-n_B(x)$, the emission rate \eq{f5} is positive for both $\omega >m\Omega$ and $\omega<m\Omega$. We observe that $\im\Pi_T(\omega-m\Omega,\omega)$ is already finite  at one-loop order because, unlike in a non-rotating plasma, where a photon cannot decay into a $q\bar q$ pair, it can do so in a rotating plasma owing to the additional supply of energy $-m\Omega$ and momentum at the boundary. Thus, from the  perspective of perturbation theory, the leading contribution to \eq{f5} carries one fewer power of  $\alpha$ (or $\alpha_s$ in quark-gluon plasma) than in the non-rotating case. 

The explicit form of $\im \Pi_T(\omega,\b k)$ at one-loop order is given in \cite{Scherer:2024uui}. At vanishing chemical potential and quark masses, it reads:
\bal
    \im \Pi_T(\omega,\b k) =& -\frac{1}{16\pi \mg k} \Bigg[ 
    \int_{\frac{\omega-\mg k}{2}}^{\frac{\omega+\mg k}{2}} d{E_p}  |\mathcal{M}_{-}|^2  (1-2n_F(E_p))\Theta(k^2)\Theta(\omega)
    + \int_{\frac{\mg k-\omega}{2}}^{\infty} d {E_p}  |\mathcal{M}_{+}|^2  2n_F(E_p)\Theta(-k^2)\nonumber\\
    + &\int_{\frac{\omega+\mg k}{2}}^{\infty} d {E_p}  |\mathcal{M}_{-}|^2  2n_F(E_p)\Theta(-k^2)+\int_{-\frac{\omega+\mg k}{2}}^{\frac{-\omega+\mg k}{2}} d {E_p}  |\mathcal{M}_{+}|^2  (1-2n_F(E_p))\Theta(k^2)\Theta(-\omega)
    \Bigg]\,, \label{o1}
\gal
where $n_F(E)= (e^{\beta E}+1)^{-1}$ is the Fermi distribution, $k^2= \omega^2-\b k^2$, and 
\ball{o4}
     |\mathcal{M}_\sigma|^2 &=  -\frac{4 \pi \alpha \sigma k^2}{\b k^2}\left[(\omega+2\sigma E_p)^2+\b k^2\right] \,.
\gal
with $\sigma=\pm 1$. The integration over $E_p$ can be performed exactly in terms of elementary and polylogarithm functions.  In the quark-gluon plasma,  one needs to replace $e^2\to e^2N_c \sum_f q_f^2 = 5e^2/3$ for two flavors.


The photon distribution in the $zx$-plane is not uniform, as is evident from its explicit dependence on $k_\bot$ and $k_z$. Let $\theta$ be the polar angle, so that $k_z= \omega\cos\theta$ and $k_\bot= \omega\sin\theta$. The observables relevant for experiment are the photon radiation rate averaged over the solid angle $do =2\pi\sin\theta d\theta$:
\ball{f7}
W(\omega)=\frac{1}{4\pi}\int  \frac{d\dot w_\text{em}}{2\pi k_\bot dk_\bot dk_z V}\bigg|_{\substack{k_z= \omega\cos\theta \\ k_\bot= \omega\sin\theta}}\,do\,,
\gal
and the elliptic flow
\ball{f20}
v_2=\aver{\cos\left[ 2\left(\frac{\pi}{2}-\theta\right)\right]}= -\aver{\cos 2\theta}\,,
\gal
where the angular brackets denote the average over the solid angle $do$ with the weight \eq{f5}. The $m=0$ contribution to the spectrum behaves as $1/k_\bot$ at small transverse momenta. This causes a logarithmic divergence when integrated over $\theta$. It is regulated by imposing a cutoff  $k_\bot>1/\mathcal{R}$.

\begin{figure}[ht]
\begin{tabular}{cc}
      \includegraphics[width=0.45\linewidth]{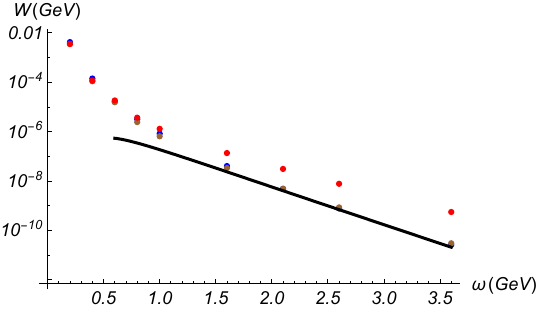} &
      \includegraphics[width=0.45\linewidth]{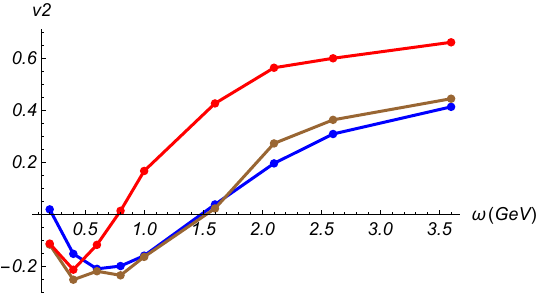}
      \end{tabular}
  \caption{Photon spectrum $W$ (left panel) and its elliptic flow $v_2$ (right panel) emitted by a rotating quark-gluon plasma at temperature $T=0.3$~GeV versus photon's energy. Blue dots: $\mathcal{R}=2$~fm, $\Omega=0.05$~fm$^{-1}$, brown dots: $\mathcal{R}=3$~fm, $\Omega=0.05$~fm$^{-1}$, red dots: $\mathcal{R}=3$~fm, $\Omega=0.25$~fm$^{-1}$, black line: leading order perturbative calculation for non-rotating plasma \eq{z7}. }
\label{fig:spect}
\end{figure}

\fig{fig:spect} exhibits the spectrum of photons emitted by a rotating plasma, assuming  the rotating region has a cylindrical shape. The spectrum and elliptic flow do not show significant dependence on the radial plasma size $\mathcal{R}$, which was taken to be 2 and 3 fm. The rotation velocity $\Omega=0.05$~fm$^{-1}$ is inferred from hydrodynamic simulations \cite{Csernai:2013bqa,Becattini:2015ska,Deng:2016gyh,Jiang:2016woz,Xia:2018tes}. A larger value of $\Omega=0.25$~fm$^{-1}$ was studied to determine the sensitivity of the spectrum and elliptic flow. The result for the spectrum is contrasted with the leading perturbative calculation for a non-rotating plasma \eq{z7}, shown as the solid black line. It is observed that at and above 1.5~GeV, the rotating and non-rotating calculations are of similar magnitude. However, at lower photon energies, the number of photons emitted by the rotating plasma significantly exceeds that of the non-rotating one. 

As mentioned in the previous section,  only those modes with $|m|\lesssim\omega \mathcal{R}$  contribute significantly to total photon yield. At low $\omega$, therefore, only the lowest $m$ must be taken into account. This effect is clearly visible in \fig{fig:super}. The spectrum of these infrared modes scales with the emission angle as $1/\theta$, as implied by \eq{d14}. This indicates that photons are predominantly emitted at small angles $\theta$, causing $v_2$ to become negative, which is consistent with \fig{fig:spect}. Note that $v_2$ is not an additive quantity. Therefore, the result displayed in \fig{fig:spect} cannot be  compared directly with experiment, but must be weighed against other contributions. Nevertheless, \fig{fig:spect} illustrates the expected trend. 

The superradiant modes' contribution to the total spectrum displayed in \fig{fig:spect} is negligible. For example, $W_\text{super}\sim 10^{-21}$~GeV at $\omega=0.1$~GeV.


\section{Superradiant amplification  of the classical magnetic field}\label{sec:g}

The quark-gluon plasma produced in relativistic heavy-ion collisions is magnetized \cite{Kharzeev:2007jp,Skokov:2009qp,Voronyuk:2011jd,Bzdak:2011yy,Bloczynski:2012en,Deng:2012pc,Tuchin:2013apa,Zakharov:2014dia}. The magnetic field dissipates over the course of the plasma's lifetime due to various inelastic processes. However, as was first pointed out by Zeldovich \cite{Zeldovich:1971ffh}, this holds only when $\omega>m\Omega$. Modes satisfying the superradiance condition $\omega<m\Omega$ are instead enhanced in a rotating plasma, according to \eq{d13}. 

Let $N(k_z,k_\bot,m;t)$ denote the number of photons in the magnetic field at time $t$ at a given $k_z$, $k_\bot$ and $m$. The effect of plasma rotation is to change this number of photons \cite{Endlich:2016jgc}:
\ball{g1}
\Delta N (k_z,k_\bot,m;t)= P(k_z,k_\bot,m)N(k_z,k_\bot,m;t)\,,
\gal
where $P(k_z,k_\bot,m)$ is the difference between the emission and absorption probabilities. Define the partial rate 
$\gamma (k_z,k_\bot,m)$ as the summand in \eq{d13}:
\ball{g2}
\frac{d\dot w}{2\pi k_\bot dk_\bot dk_z V}=\sum_{m=-\infty}^\infty \gamma (k_z,k_\bot,m)\,.
\gal
Then, $P(k_z,k_\bot,m)= \gamma (k_z,k_\bot,m)\Delta t$. The change in the number of photons in the magnetic field therefore obeys the equation
\ball{g3}
\frac{dN(k_z,k_\bot,m;t)}{dt}=  \gamma (k_z,k_\bot,m)N(k_z,k_\bot,m;t)\,.
\gal
As mentioned, $\gamma (k_z,k_\bot,m)>0$ when $\omega<m\Omega$ and is negative otherwise. As a result, the superradiant modes grow  exponentially with time.

\begin{figure}[ht]
\begin{tabular}{cc}
      \includegraphics[width=0.45\linewidth]{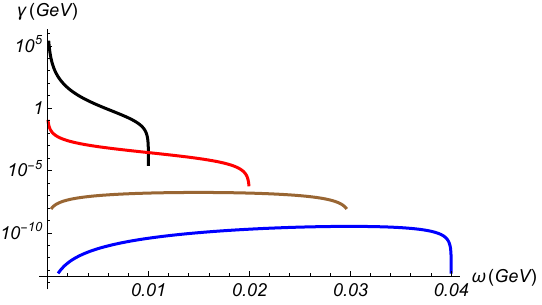} &
      \includegraphics[width=0.45\linewidth]{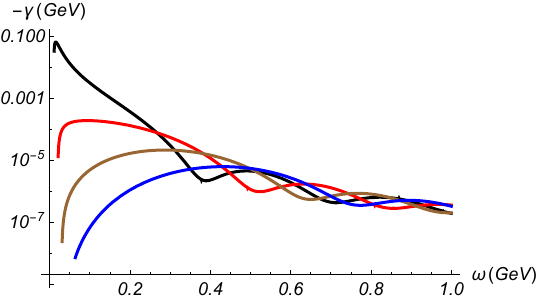}
      \end{tabular}
  \caption{The growth (left panel, $\omega<m\Omega$) and damping (right panel, $\omega>m\Omega$) rates versus photon's energy in rotating quark-gluon plasma with $T=0.3$~GeV, $\mathcal{R}=2$~fm, $\Omega=0.01$~GeV at $\theta=\pi/2$ (in the `reaction plane'). Black: $m=1$, red: $m=2$, brown: $m=3$, blue: $m=4$. Notice that plots have difference ranges. 
}
\label{fig:super}
\end{figure}

The growth rates of the different $m$-modes are displayed in \fig{fig:super}. The growth $\gamma$ is positive for the superradiant modes $\omega<m\Omega$ and negative otherwise; in the latter case, $-\gamma$ represents the damping rate.  The exponential growth of the photon number indicates an instability of the magnetic field in the rotating plasma. This instability converts the plasma's energy into infrared magnetic-field modes. Given that the quark-gluon plasma's lifetime is about 10~fm, only the instability of $m=1,2$ modes is relevant for phenomenology. This instability might manifest itself as a spike of photons with $\omega\lesssim 10$~MeV observed in the detector. 

The damping rates shown in the right panel of \fig{fig:super} are small and have little phenomenological significance for $\omega>0.2$~GeV.

\section{Summary}\label{sec:k}

The main result of this paper is the differential rates of photon emission and absorption by a uniformly rotating plasma. The differential emission rate is given by \eq{f5}, and the difference between the emission and absorption rates by \eq{d13}. These rates are already finite at one loop order, in contrast to the non-rotating case, where the leading contribution starts at two-loop order. Furthermore, photons with $\omega<m\Omega$ exhibit superradiance: their emission rate exceeds their absorption rate. 

Using the known expressions for the spectral density of the current-current correlator at one-loop, the differential rate of photon emission from the quark-gluon plasma was obtained. The angular average of the photon rate $W$ and its elliptic flow $v_2$ are displayed in \fig{fig:spect} as a function of photon energy. At large  $\omega$, the calculated rate is comparable to the leading perturbative result for non-rotating plasma; however, in the infrared region $\omega<1$~GeV it significantly exceeds it. The elliptic flow tends to be large and positive above 1.5~GeV, and negative in the infrared. These results  suggest that a solution to the  direct photon puzzle \cite{David:2019wpt, Gale:2021emg, Paquet:2015lta} may lie in taking quark-gluon plasma rotation into account.

The quark-gluon plasma  produced in relativistic heavy-ion collisions is accompanied by a strong magnetic field. Although the field strength drops by several orders of magnitude from its peak value, it remains fairly stable through most of the plasma's lifetime \cite{Stewart:2021mjz}. Its absorption rate is quite low for most of the photon modes, as  seen in the right panel of \fig{fig:super}. However, the superradiant modes with $\omega<m \Omega$ are inherently unstable, since their emission rate exceeds their absorption rate. The left panel of \fig{fig:super} shows the growth rate of these superradiant modes. Given the quark-gluon plasma's lifetime of about 10~fm, the modes $m=1,2$ can significantly amplify the magnetic field in the deep infrared region $\omega<2\Omega\lesssim 10$~MeV. This could plausibly produce a spike of photons at these energies in detectors.

It has been extensively discussed in the literature that a chiral plasma can convert part of its energy and chirality into magnetic field by means of the chiral magnetic instability \cite{Carroll:1989vb,Joyce:1997uy,Boyarsky:2011uy,Kharzeev:2013ffa,Khaidukov:2013sja,Kirilin:2013fqa,Akamatsu:2013pjd,Avdoshkin:2014gpa,Dvornikov:2014uza,Tuchin:2014iua,Manuel:2015zpa,Buividovich:2015jfa,Sigl:2015xva,Xia:2016any,Kaplan:2016drz,Kirilin:2017tdh,Tuchin:2018sqe,Mace:2019cqo}. However, this instability develops over timescales exceeding the quark-gluon plasma's lifetime by an order of magnitude, and therefore has no phenomenological relevance for relativistic heavy-ion collisions. A tantalizing possibility is that the chiral magnetic instability and the superradiance-driven instability discussed in \sec{sec:g} mutually reinforce each other's growth. This is left as an avenue for future research.

The problem of magnetic-field instability in a rotating plasma was previously examined in \cite{Das:2025kgq},  which concluded that although rotation renders the magnetic field unstable, the slow growth of the unstable modes makes this instability irrelevant for the quark-gluon plasma phenomenology. The phenomenon of superradiance overturns this conclusion.

A cylinder of radius $\mathcal{R}$ rigidly rotating with constant angular velocity $\Omega$ about its symmetry axis was used to model the region of rotating plasma, and the non-inertial forces associated with rotation were neglected.  The parameters $\mathcal{R}$ and $\Omega$ are inferred from the vorticity distribution in hydrodynamic simulations of the quark-gluon plasma \cite{Csernai:2013bqa,Becattini:2015ska,Deng:2016gyh,Jiang:2016woz,Xia:2018tes}. Fortunately, the photon spectrum shows little dependence on $\mathcal{R}$; the cylinder's height, moreover, plays no role at all, as it cancels out of the equations. Although only a single rotating cylinder was considered here, a more realistic description may require an ensemble of cylinders rotating with different angular velocities. The effect of the non-inertial forces on photon production deserves a dedicated study. 

For the spectral density, we used the one-loop expression in the chiral limit. A more refined treatment would employ resummed fermion propagators; this is expected to moderately reduce the photon rate, though the qualitative picture should remain unchanged. The impact on $v_2$ is more difficult to assess, given its sensitivity to other model parameters. These and other issues relevant to phenomenological applications warrant further analysis.

\acknowledgments
This work  was supported in part by the U.S. Department of Energy under Grant No.\ DE-SC0023692.

\appendix
\section{Non-rotating limit}\label{app:A}

In the particular case of a non-rotating plasma, the rotation matrix $R^{ij}$ is replaced by the unit matrix $\delta^{ij}$. Assuming the photon is in a state of given momentum $\b k$ and energy $\omega= |\b k|$, the photon wave function is given by
\ball{z1}
\langle 0| A^k(x)|\gamma(\b k,\lambda)\rangle = \frac{1}{\sqrt{2\omega V}}\b\epsilon_{\b k, \lambda}e^{-i\omega t+i\b k\cdot \b x}\,,
\gal
where $\lambda$ denotes the photon's polarization. Substituting this into \eq{b11},\eq{b15}, and \eq{b20}, using the polarization sum 
\ball{z3}
\sum_\lambda \epsilon^i_{\b k, \lambda}\epsilon^{*j}_{\b k, \lambda}= \delta^{ij}-\frac{k^ik^j}{\b k^2}\,,
\gal
noting that the tensor $\Pi^{ij}(k)$ is transverse, and recalling $d\Gamma = d^3kV/(2\pi)^3$, one derives the rate of photon emission per unit volume:
\ball{z5}
\frac{d\dot w}{V}\bigg|_\text{no rotation}= \frac{1}{\omega}\im\Pi_R^{ii}(k) \frac{d^3k}{(2\pi)^3}\,.
\gal
 This is a well-known result \cite{Weldon:1983jn,McLerran:1984ay,Gale:1990pn}. In the non-rotating plasma, the first non-vanishing contribution to the spectral density arises at two-loop order in perturbation theory, due to gluon exchange. The result is
\ball{z7}
 \frac{d\dot w}{d^3k V}\bigg|_\text{no rotation}= \frac{5}{9}\frac{\alpha\alpha_s}{2\pi^2\omega}T^2 e^{-\omega/T}\ln\left(\frac{2.9\omega}{g^2 T}\right)\,.
\gal

\bibliography{/Users/tuchin/Articles/references}

\end{document}